\documentclass[fleqn,10pt]{wlscirep}
\usepackage{tabularx}
\usepackage{caption}
\usepackage{float}
\usepackage{color}
\usepackage{textcomp}
\title{Quantifying birefringence in the bovine model of early osteoarthritis using polarisation-sensitive optical coherence tomography and mechanical indentation}

\author[1,*]{Matthew Goodwin}
\author[1]{Bastian Br\"{a}uer}
\author[2]{Stephen Lewis}
\author[2]{Ashvin Thambyah}
\author[1]{Fr\'{e}d\'{e}rique Vanholsbeeck}
\affil[1]{The Dodd-Walls Centre for Photonic and Quantum Technologies, Department of Physics, The University of Auckland, Private Bag 92019, Auckland, New Zealand.}
\affil[2]{Department of Chemical and Materials Engineering, The University of Auckland. Private Bag 92019, Auckland, New Zealand.}

\affil[*]{mbro632@aucklanduni.ac.nz}


\begin{abstract}
Recent studies have shown potential for using polarisation sensitive optical coherence tomography (PS-OCT) to study cartilage morphology, and to be potentially used as an \textit{in vivo}, non-invasive tool for detecting osteoarthritic changes. However, there has been relatively limited ability of this method to quantify the subtle changes that occur in the early stages of cartilage degeneration. An established mechanical indenting technique that has previously been used to examine the microstructural response of articular cartilage was employed to fix the bovine samples in an indented state. The samples were subject to creep loading with a constant compressive stress of 4.5 MPa and, when imaged using PS-OCT, enabled birefringent banding patterns to be observed. The magnitude of the birefringence was quantified using the birefringence coefficient (BRC) and statistical analysis revealed that PS-OCT is able to detect and quantify significant changes between healthy and early osteoarthritic cartilage ($p<0.001$). This presents a novel utilization of PS-OCT for future development as an \textit{in vivo} assessment tool.
\end{abstract}
\begin{document}
\flushbottom
\maketitle
%
%
\thispagestyle{empty}

\section*{Introduction}

Osteoarthritis (OA) is a debilitating disease that affects articulating joints, in which there is a 	gradual wearing away of the aneural cartilage covering the bone ends, resulting in pain. A standard 	clinical diagnosis of OA is  made via weight bearing X-rays, in which joint space narrowing is seen as the hallmark sign of cartilage loss.\cite{Diag1,Diag2} Cartilage being aneural typically means that diagnosis of OA is often too late, and for the patient the options frequently are standard analgesics, or when that is ineffective, total joint replacement surgery.\cite{Pain1,Knee1} There is no cure for OA, and prevention strategies are highly limited. Many believe one of the major obstacles in OA research is in determining the early or even pre-OA state of the joint. Such a joint would appear to have normal full thickness cartilage but changes at the micro-to-nano scales. In the very early stages of cartilage degeneration, proteoglycan depletion coupled with alterations in the collagen fibril network results in swelling and increased tissue permeability.\cite{depthwise} The intrinsic repair capability of articular cartilage is highly limited and, once exhausted, destruction of the extracellular matrix (ECM) ensues.\cite{morpho} ECM destruction is one of the first macroscopic signs of OA and articular cartilage (AC) surface integrity is considered one of the key biomarkers to quantify OA progression.\cite{detectearlyoa} The challenge for most OA researchers, especially in the imaging domain, would be to develop ways in which this early and very mild cartilage degeneration is detected.\\
\indent
To this end, two recent developments have arisen that also provide the motivation for this study. First, is the recent validation of an animal model for early OA, in which bovine patella cartilage tissue with mild degeneration has been studied extensively both structurally and mechanically.\cite{1,2} These studies have established that the collagen matrix fibrillar-scale `destructuring' takes place as an initial stage of the degenerative process.\\
\indent
The second development is the emergence of Optical Coherence Tomography (OCT) as a real-time, non-invasive, high resolution imaging technique for articular cartilage structure and health. Similar to ultrasound, conventional OCT images are created by detecting the intensity of backscattered light. Many studies have been published using OCT to evaluate cartilage degeneration and while most show promising correlation with an established cartilage assessment scheme, the lack of accurate parameters and issues regarding reliability means the   indisputable differentiation between healthy and early degenerative cartilage remains elusive.\cite{COCT1,COCT2,COCT3,COCT4,4}\\
\indent
Polarisation-sensitive OCT (PS-OCT) is a development of OCT where the polarisation state of the backscattered light is measured, enabling the detection of tissue birefringence. Tissue birefringence is dependent on collagen fibre organisation and orientation. Drexler et al. linked the loss of birefringence with the early stages of cartilage degeneration using polarised light microscopy.\cite{drexler} Other groups built upon this research and confirmed the findings with OCT that a decrease in birefringence is associated with increasing cartilage degeneration.\cite{probire1,8,probire2} A recent study by Brill et al. paramterised and quantified human cartilage degeneration using PS-OCT.\cite{3} While the results demonstrated that PS-OCT parameters may be of some additional diagnostic assistance, the most significant parameters were those derived from the intensity based OCT data. 
The authors noted that the majority of individual PS-OCT parameters such as birefringence or banding characteristics seemed to be unrelated to the samples' degree of degeneration. However, this study had two important limitations, the first was the small sample size (13 osteochondral samples from four patients) and the second was that the samples were sourced from patients who underwent total knee replacement surgery due to severe OA. These issues we believe may be overcome by the use of the above-mentioned bovine model of early OA, which provides a virtually limitless supply of cartilage tissue with mild degeneration and, with it, well established micro-to-fibrillar level structural changes. While histology is considered the gold standard of cartilage degeneration assessment, many studies have incorporated techniques that enable the determination of biomechanical properties as this may allow a more comprehensive assessment of OA.\cite{BioMech1,BioMech2,BioMech3,BioMech4} Scanning electron micrographs of mechanically indented cartilage have revealed that during early cartilage degeneration, the collagen network is disrupted and the fibres begin to aggregate leading to a reduction in structural integrity of the ECM.\cite{2} We propose that by studying the cartilage tissue in its mechanical compressed state, any alteration of the cartilage microstructure will become detectable by PS-OCT.\\
\indent
Therefore, in this exploratory study, OCT and mechanical indentation are used to detect signal differences between healthy bovine cartilage and `healthy-appearing' cartilage with known fibrillar-scale de-structuring associated with early stages of OA.\\

\section*{Results}

\subsection*{Phase evaluation}
\begin{figure}[ht]
    \centering
    \includegraphics[width=0.9\linewidth]{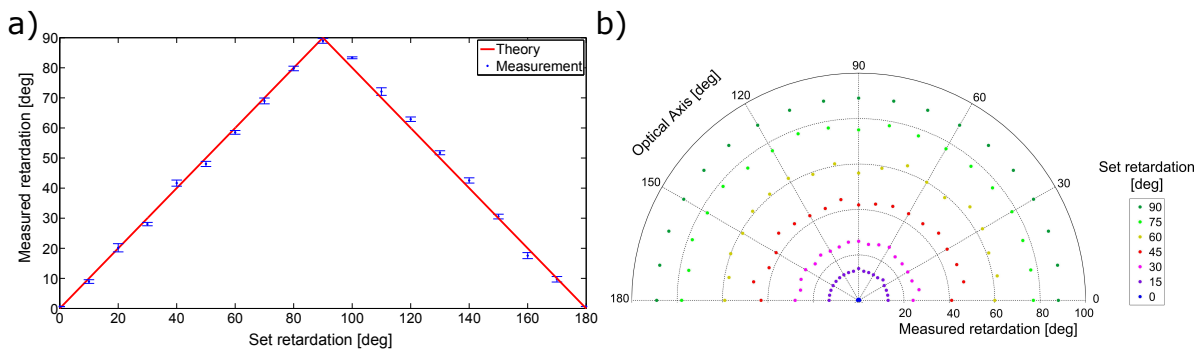}   
    \caption[Plot of the calibration measurements for the retardation]{a) Plot of measured retardation (data points) with standard deviation and set retardation for an optical axis of $20^\circ$, b) polar plot of measured retardation versus set retardation for fixed fast axis orientations. The circumference of the half circle indicates the optical axis.}
    \label{fig:Ret_OA_plot}
\end{figure}

The performance of the system was evaluated using a Berek compensator to measure known polarisation states. The compensator allows the operator to set, both, the optical axis $\phi$ and retardation $\theta$ independently of each other. First, the optic axis was fixed and the retardation was varied from  $0^\circ$ to $180^\circ$ in increments of $10^\circ$. Figure~\ref{fig:Ret_OA_plot} a) shows a plot of measured retardation angles versus set retardation at the Berek. The measured values are in good agreement with the theory (solid line) with a maximum standard deviation of $~1.4^\circ$ at $\phi=20^\circ$ set optical axis. Five repeated measurements were taken with a mean standard deviation of $~0.7^\circ$. Second, we measured the retardation for a range of optical axis orientations varied from $0^\circ$ to $180^\circ$ in increments of $10^\circ$, while the retardation was varied from  $0^\circ$ to $90^\circ$ in  $15^\circ$ increments. Again, the measured set of values agree well with a mean standard deviation of $1.6^\circ$ (figure\ref{fig:Ret_OA_plot}.b)\\
\subsection*{Image phase analysis}
\begin{figure}[htbp]
 	 \centering
  	\includegraphics[width=\linewidth]{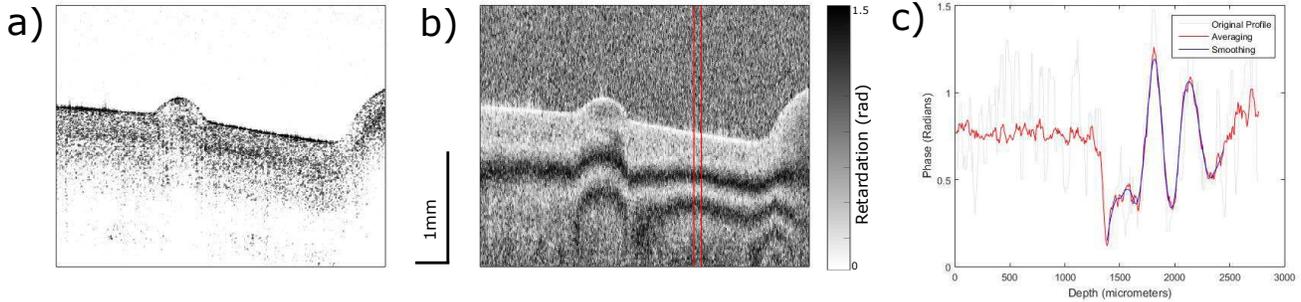}
	\caption{a) Intensity image of an indented healthy (G0) cartilage sample. b) Phase image of the same sample with the averaging window indicated by the red lines. c) Phase profile showing the original, averaged, and smoothed data.}
	\label{fig: Analysis}
\end{figure}

\begin{figure}[htbp]
  \centering
  \includegraphics[width=0.98\linewidth]{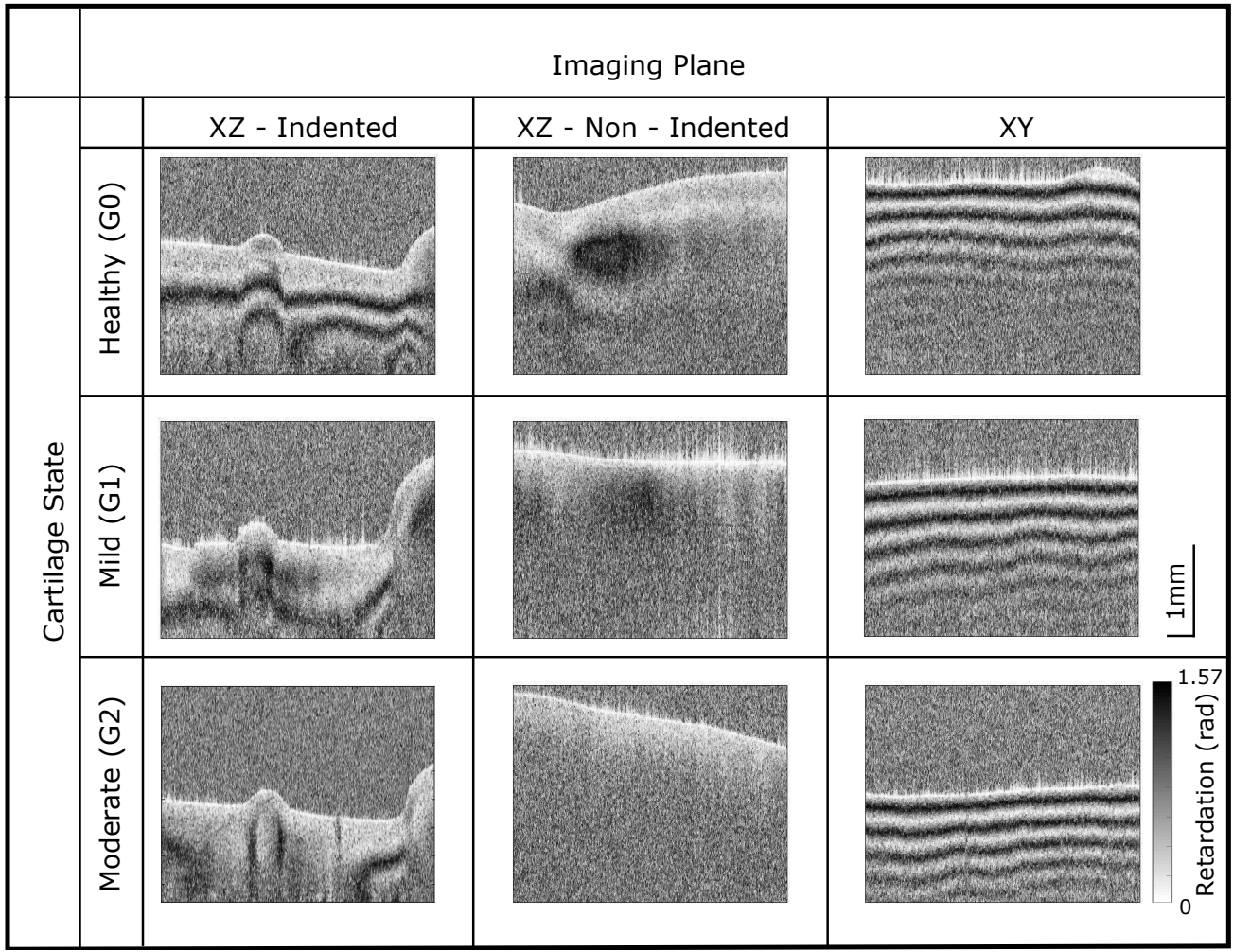}
\caption{Retardation image obtained from our PS-OCT system images showing the characteristic patterns observed. The healthy cartilage (G0) exhibits strong optical activity when imaged in the indented section and with increasing degeneration, the optical activity decreases. Colour and scale bar correspond to all images.}
\label{fig:retimage1}
\end{figure}

Out of the 26 samples, 10 were healthy (G0), 8 had mild degeneration (G1), and 8 had moderate degeneration (G2). All samples underwent compression using the indentation device. Since the PS-OCT setup detects orthogonal phase amplitudes we can obtain intensity (figure~\ref{fig: Analysis}.a) and phase images (figure~\ref{fig: Analysis}.b). The system has a $\frac{\pi}{2}$ phase wrap, the retardation gradient was calculated as the cumulative, absolute phase change divided by the depth of the smoothed profile (see method) (figure~\ref{fig: Analysis}.c). PS-OCT images were taken in three different locations on the cartilage corresponding to three areas of interest (figure~\ref{fig:retimage1}). \\

\begin{figure}[htbp]
 	 \centering
 	 \includegraphics[width=\linewidth]{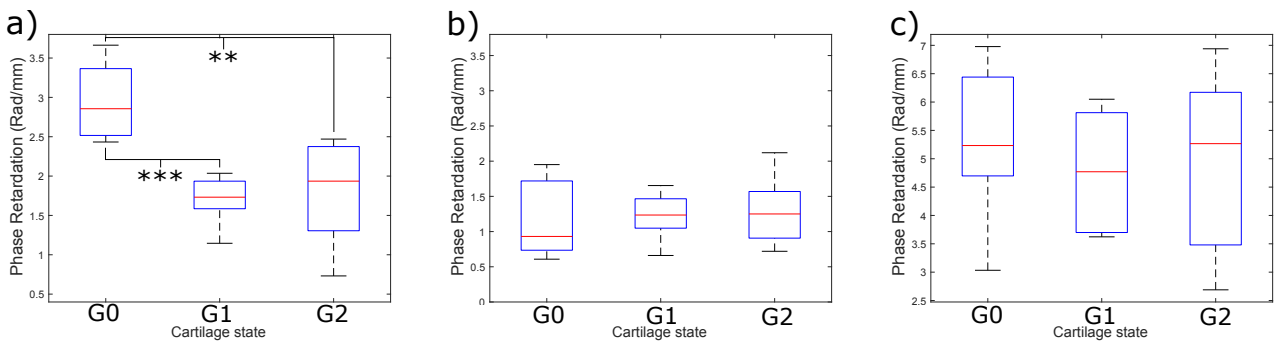}
	\caption{a) Boxplot showing results from phase analysis for PS-OCT images taken in the XZ - indented region. Results show there is very strong statistical significance between healthy (G0) and mildly degenerate cartilage (G1) (***$p~value < 0.001$) and a strong significance between healthy and moderately degenerate (G2) cartilage (**$p~value < 0.01$). b) Boxplot showing results from phase analysis for  images taken in the XZ - non indented region. c) Boxplot showing results from phase analysis for XY plane images. No significant relationship exists between healthy and OA states for non-indented and XY plane images.}
	\label{fig: boxplot}
\end{figure}

From the images of the indented region (Figure~\ref{fig:retimage1}) and the data from table~\ref{table:result}, it appears that there is a reduction in birefringence with increasing degeneration. Calculating the retardation gradient has confirmed that a statistically significant (p $<$ 0.001) relationship between healthy (G0) and early OA (G1 and G2) indented cartilage exists (figure~\ref{fig: boxplot}.a) with the mean retardation gradient for healthy of 3.0 $\pm$ 0.5 radians per millimeter. Mild and moderate OA degeneration had a mean retardation gradient of 1.7 $\pm$ 0.3 and 1.8 $\pm$ 0.7 radians per millimeter respectively (table~\ref{table:result}). Whereas, images obtained from imaging in the non-indented region in the XZ plane and images in the XY plane show very little difference with increasing degeneration (figure~\ref{fig:retimage1}).  Phase analysis showed that there is no statistical significance between any of the three groups investigated (figure~\ref{fig: boxplot}b \& c).\\

\begin{table}[htbp]
\centering
\caption{\bf Summary of results for phase analysis calculating the retardation gradient in various imaging planes.}

\begin{tabular}{ccccc}
\hline
\hline
 & & XZ - Indented & XZ - Non-indented & XY \\
\hline
G0& & 3.0 $\pm$ 0.5& 1.1 $\pm$ 0.5 &5.2$\pm$ 1.0 \\
G1& &   1.7 $\pm$ 0.4 & 1.2 $\pm$ 0.3 & 4.9 $\pm$ 1.0\\
G2& & 1.8 $\pm$ 0.7& 1.3 $\pm$ 0.5 & 5.4 $\pm$ 1.5\\
\hline
Anova& P-value& $<$ 0.001& 0.770& 0.693\\
\textit{Post-hoc} testing& G0 vs. G1 & *** & ns & ns\\
 & G0 vs. G2 & ** & ns & ns \\
 & G1 vs. G2 & ns & ns & ns \\
\hline
\hline
\end{tabular}\\
\vspace{1mm}
Values are given in radians per millimeter and expressed in the form: mean $\pm$ standard deviation.\\
Levels of significance are indicated by: ***\textit{p}-values of \textit{p} $\leq$ 0.001, **\textit{p}-values of 0.001 $<$ \textit{p} $\leq$ 0.01, ns - not significant
  \label{table:result}
\end{table}

\section*{Discussion}
The most important findings of this study is that the results agree with literature that previously stated there is little relationship between the optical properties of cartilage and the degree of degeneration without the use of indentation. Coupling PS-OCT with mechanical loading has shown that there is a significant relationship between the observed birefringence and the, subjectively graded, health state of the cartilage. Previous studies that have examined OA degeneration in cartilage have discussed using the retardation gradient as a parameter to quantifying cartilage degeneration, but none have been successful in finding a method that produces statistically significant results.\cite{3,6}\\
As mentioned above, one of the key points of conflict in previous literature is whether or not cartilage degeneration affects the optical activity observed using PS-OCT. Xie et al. investigated the birefringent characteristics associated with cartilage degeneration and found that bovine samples exhibit very little polarisation sensitivity in both healthy and early degenerate samples.\cite{Xie06} The authors' noted that only in the later stages of degeneration were birefringent banding patterns present and could be used to differentiate severe degeneration from early degenerative and healthy samples. Furthermore, Xie et al. found topographical variations of polarisation sensitivity exist within the bovine articular joint found that the presence or absence of optical activity was unrelated to the degree of degeneration.\cite{Xie08} However, other studies have established the opposite; banding patterns are present with healthy samples and are lost with increasing cartilage degeneration.\cite{probire2,probire1,Herr1} As a result, the specific methodology and intra-cartilage variation may have lead to misinterpretation of characteristic birefringent patterns with increasing degeneration.\\
Ugryumova et al. identified that as equine cartilage is rotated in the sagittal plane, banding patterns can be observed due to the deep zone fibre alignment no longer being parallel with the incident light.\cite{5} Simiarly, Xie et al. demonstrated that bovine cartilage shows very little polarisation sensitivity when imaged in the XZ plane but when rotated to image the XY plane, a strong banding pattern was present due to the highly aligned deep zone fibres.\cite{6} The  believed reason for the apparent non-birefringence of the sample when scanned in the XZ plane is that the incident light is parallel to the deep-zone fibre orientation. Therefore the light travels through as if the cartilage is isotropic.\cite{5} During mechanical loading of healthy cartilage, the collagen fibre alignment becomes non parallel in relation to the optical axis leading to an observable birefringent signal. It is believed that with degenerative samples the loss of alignment and/or decreased integrity of the collagen matrix results in a non-uniform deformation of the collagen fibres leading to a decrease in the birefringence observed when indented.\\
\indent
In line with our results, several studies that also use cartilage samples from large animals (bovine, equine) fail to show optical activity when imaging normal to the surface (XZ plane) without any form of indentation.\cite{5,6,Xie06,Xie08} In contrast,  studies using healthy human samples have observed strong birefringent patterns.\cite{probire1,probire2} It remains unclear why banding patterns are seen in non-modified human cartilage yet absent in bovine. Jahr et al. suggested these discrepancies may be due to species specific differences.\cite{detectearlyoa} In support of this view, Pederson et al. found that the superficial zone of human articular cartilage is thicker than that of bovine articular cartilage which may explain the differences observed.\cite{HumanBovine}. It would be beneficial for future experiments to determine any correlation between bovine and human cartilage using PS-OCT and mechanical indentation.\\
\indent
One of the main limitations of our study is the analysis process requires user input. The indented XZ images shown in figure~\ref{fig:retimage1} have asymmetrical banding patterns around the center bulge and hence the A-scan selection will influence the magnitude of the BRC. In order to minimize this, the same approximate location on each image was used for analysis. Furthermore, until the differences between human and bovine cartilage have been reconciled, the transferability of the results remains limited.\\
\indent
In conclusion, this study demonstrates that differentiation of healthy from early degenerative cartilage using PS-OCT is possible. Using mechanical indentation, optical activity can be induced in cartilage samples that would otherwise exhibit anisotropy. The magnitude of the induced birefringence is related to the samples degree of degeneration, allowing clinically useful information to be obtained. Our study provides the basis for future studies to build upon and determine the diagnostic value of this technique.
\section*{Methods}

\subsection*{Polarisation-Sensitive Optical Coherence Tomography}
\begin{figure}[htbp]
  \centering
  \includegraphics[width=0.95\linewidth]{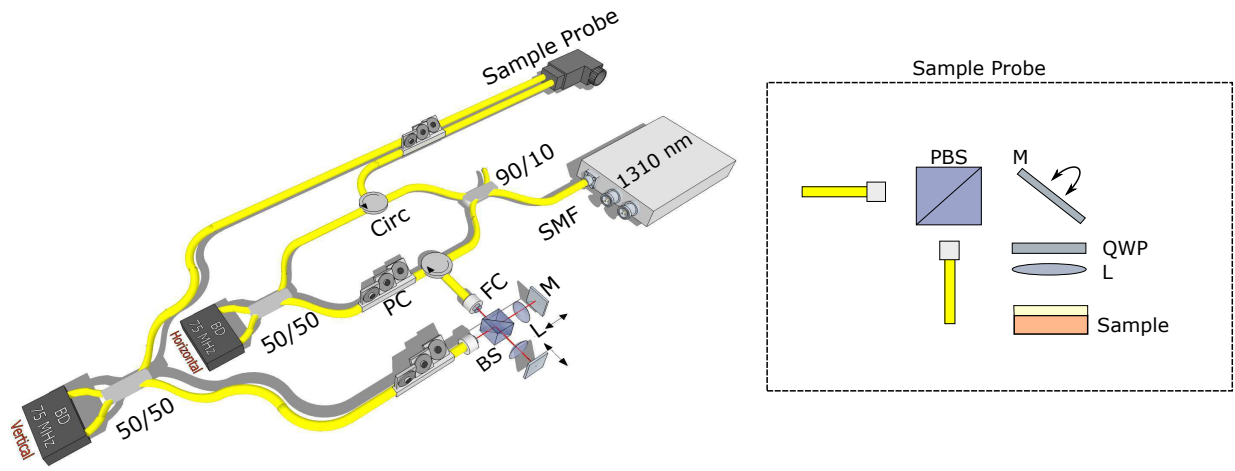}
	\caption{Diagram of the PS-OCT setup used. 1310: 1310nm light source, SMF: Single Mode Fibre, 90/10: 90/10 Fibre coupler, Circ: Circulator, 50/50: 50/50 Fibre coupler, FC: Collimator, PC: Polarisation Controller, BS: 50/50 Beam Splitter, L: 50mm lens, M: Mirror, PBS: Polarisation Beam Splitter, QWP: Quarter Wave Plate.}
	\label{fig: Setup}
\end{figure}

Images were taken using an in-house built swept-source PS-OCT system which uses a combination of fibre components and free space optics (Figure~\ref{fig: Setup}). A commercially available swept source laser source was used (AXP50125-6, Axsun Technologies Inc, Billerica, Massachusetts) that has a central wavelength of 1310 nm, a bandwidth of 100 nm, a 50 kHz sweep rate, and has an average output power of 41 mW. The system has an  axial resolution of 10 \textmu m in air according to the spectrum, a lateral resolution of 20 \textmu m and the optical power incident on the sample was measured to be 11 mW.\\
\indent
The optical beam is split into the sample and reference arm using a 90/10 fibre optic coupler. In the sample arm, a quarter wave plate (QWP) is used to circularly polarise the incident light. This helps ensure the phase measurements are independent of sample rotation.\cite{Rot1} Upon reflection in the sample arm, the light is split into orthogonal polarisation states through the use of a polarizing beamsplitter and coupled into two single mode fibres. In the reference arm the light is split by a non-polarizing beamsplitter to enable the path length of each polarisation state to be adjusted independently. The orthogonal components from the sample interfere with one of the reference arms each. A balanced photodetector is used to detect the amplitude of the polarisation components \textit{A$_{V}$} and \textit{A$_{H}$}.

\indent
The signal obtained from the detector is then sent for data processing and the images are reconstructed in LabView software. The reflectivity, \textit{R(z)}, and retardation, $\delta$(z), can be calculated using the following formula:
\begin{equation}
R(z) \propto A_{V}^2(z) + A_{H}(z)^2
\end{equation}
\begin{equation}
\delta(z) = arctan(\frac{A_{V}(z)}{A_{H}(z)})
\end{equation}
\indent
The B-scan images were obtained using a galvoscanner setup operating at 70 Hz with a total image width of 10mm.
\subsection*{Cartilage Classification}
\begin{figure}[htbp]
  \centering
  \includegraphics[width=0.8\linewidth]{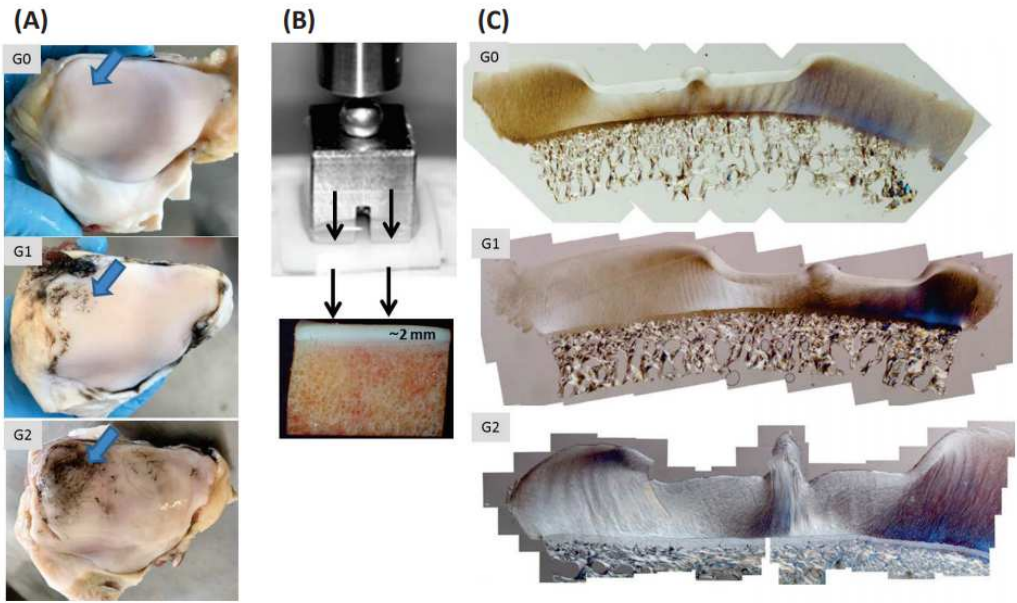}
\caption{: (A) Bovine patellae from healthy (G0) , mildly degenerate (G1) and moderately degenerate (G2 ) samples after India ink was applied. Note that the India ink washes off an intact surface, but stains positive, those regions with microscopic defects. The large arrows point to regions sampled for mechanical testing with channel indentation.  (B) Channel indentation involves a flat-ended stainless steel indenter compressed against cartilage – except there is a channel space (1 mm gap) in the indenter that allows the cartilage tissue to ‘bulge’ up into the channel space. (C) A series of differential interference contrast optical light microscopy images, of the cartilage-on-bone after indentation and formalin-fixation under load.}
\label{fig: CartPrep}
\end{figure}

The bovine patellae tissue used in this study serves as a convenient tissue source for studying early cartilage degeneration. Validated previously as a model for human early osteoarthritis,\cite{1} this large animal tissue is sourced from the local meat processing plant and stored at -20 degrees Celsius. The bovine are raised primarily for the dairy industry and are allowed to live as long as 5 to 9 years. The model allows researchers to select a range of grades of cartilage degeneration from simple visual qualitative analysis following India ink staining.\cite{1,9} Importantly, the surface staining and subsequent grading into G0, G1 and G2, as used in the present study, has previously been shown to represent deep matrix fibrillar-scale changes in the cartilage tissue.\cite{physiorange1,Indent2,2} The mild to moderate degenerative changes in the cartilage surface tends to be localised to the distal-lateral quarter of the patella and can be assessed accurately by India ink staining. Following India ink staining, patella gross appearance was categorised using the Outerbridge classification\cite{9} yielding three different groups: G0 (no stain uptake and cartilage appear completely intact), G1 (mild stain uptake with slight softening and swelling of the cartilage), and G2 (with increased stain uptake and some surface fissures). Both G1 and G2 were found to be equivalent to a clinical OOCHAS score of 1 to 4, which upon further correlation with microstructure, was defined as the mildly degenerate or early osteoarthritic state.\cite{pritzker} Using India ink staining, a total of 10 intact, 8 mildly degenerate, and 8 moderately degenerate patellae were selected (figure~\ref{fig: CartPrep}.a).  A cartilage-on-bone block of approximately 14mm x 14mm x 14mm was cut from the distal lateral quarter of each patella, the region for which patella classified as mildly degenerate contained the surface disruption. All the samples used in this study were obtained from a local meat processing plant. As such ethics approval was not required.

\subsection*{Cartilage Indentation}
\begin{figure}[htbp]
  \centering
  \includegraphics[width=0.4\linewidth]{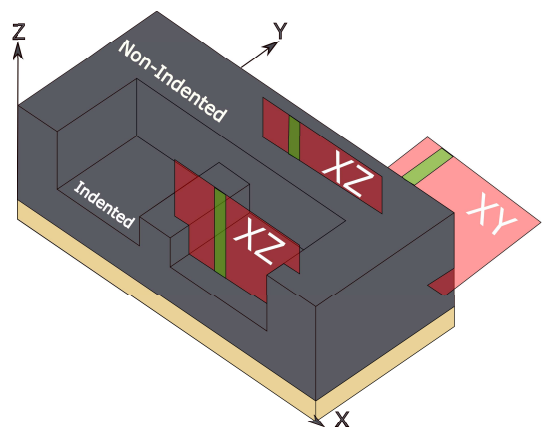}
\caption{Schematic representation of the indented cartilage samples. B-scans were taken in both the XZ and XY plane (red planes). XZ images were further classified depending on if the image was taken in the indented or non-indented region. Green planes indicate the approximate regions where gradient analysis was performed. }
\label{fig: CartilagePlane}
\end{figure}

The cartilage samples were mechanically loaded, fixed and decalcified. The loading protocol of 4.5 MPa follows that of an earlier channel-indentation.\cite{2} In that study the mean axial strains following creep loading at 4.5 MPa were recorded as 56\% (range 39\%–68\%). The loading magnitude of 4.5 MPa, arguably, is within the range of cartilage stresses derived from studies of the human knee in compression.\cite{physiorange2} The idea that the collagen network disruption in the mild to moderately degenerate tissue groups, G1 and G2, are not a result of the loading, is supported by an earlier microstructural study without mechanical loading showing that such fibrillar matrix destructuring occurs inherently in G1 and G2 tissues.\cite{Indent2} The effect of the indenter can be visualised by the schematic in figure~\ref{fig: CartilagePlane}.

\subsection*{Cartilage Preparation}
The sample blocks were frozen in 0.15M saline for transport and storage until use. Prior to imaging the samples were defrosted and allowed to equilibrate in a saline bath for up to 2 hours, each sample only underwent one freeze-thaw cycle. The samples were drained on paper to remove excess fluid.  Images were taken with the laser 90$^\circ$ to the surface (XZ plane) in the indented and non-indented regions. Images were also taken on the side of the cartilage block with the laser 0 $^\circ$ to the surface, (XY plane) as shown in figure~\ref{fig: CartilagePlane}. The total measurement time (time samples were out of the saline bath) was kept under 2 minutes.\\

\subsection*{Image Processing}
Post-image processing was carried out using custom imaging algorithms created in MATLAB. The primary parameter used to analyse the strength of the birefringent is the retardation gradient, also known as the birefringence coefficient (BRC). The user was prompted to select the approximate A-scan where the gradient analysis was performed. A averaging window of 10 A-scans (200 \textmu m window) was applied around the user selected point. The algorithm performed surface detection and a threshold was applied to determine the maximum imaging depth. Using the systems resolution in air and taking the refractive index of bovine cartilage as n=1.36\cite{RI} the depth calculation was calibrated to the tissue. A running average was applied to smooth the phase profile between the surface and the maximum depth. The retardation gradient was calculated as the cumulative, absolute change in phase of the `smoothed' data divided by its depth. This measurement was repeated across 5 adjacent B-scans (25 \textmu m separation between each B-scan) before a final value was determined for each sample.\\
\indent
Statistical analysis was completed through MATLAB. The assumptions of normality and equal variance were tested  using the \textit{vartestn} function. One way ANOVA was used to identify if any relationships existed between the three groups (G0,G1,G2) in each imaging plane using the \textit{anova1} function.  Tukey's \textit{post-hoc} test was performed using the \textit{multcompare} function to quantify the significance of the relationship. P-values $\leq$0.05 were considered statistically significant.\\

\subsection*{Data Availability}
The datasets generated and analysed are available from the corresponding
author on reasonable request.
\bibliography{Ref}

\section*{Acknowledgements}

The authors would like to acknowledge funding from Marsden Fund and Royal Society of New Zealand (UoA1509) which made this research possible.

\section*{Author contributions statement}

M.G Analysed the data, M.G  and S.L conducted the experiments, B.B calibrated the imaging system, A.T prepared the cartilage samples, F.V conceived the experiments and obtained funding for the research. A.T and F.V supervised the research. All authors interpreted the results and proofread the manuscript.

\section*{Additional information}
\subsection*{Competing interests}
The authors declare no competing interests.

\end{document}